\documentclass[aps,prl,twocolumn,preprintnumbers,superscriptaddress]{revtex4}
\usepackage[colorlinks,linkcolor=blue,citecolor=blue]{hyperref}
\usepackage{amsmath}    
\usepackage{amssymb}
\usepackage{graphicx}
\usepackage{amsfonts}   
\usepackage{latexsym}   
\usepackage{slashed}
\usepackage{yfonts} 
\usepackage{verbatim}
\usepackage{wrapfig}

\let\non\nonumber

\global\newcount\itemno \global\itemno=0

\def\itemaut#1{\global\advance\itemno by1\noindent\item{\the\itemno.}#1}

\newif{\ifeq}           
\eqtrue                 

\newcommand{\be}{\begin{equation}}
\newcommand{\ee}{\end{equation}}
\newcommand{\bes}{\begin{equation*}}
\newcommand{\ees}{\end{equation*}}
\newcommand{\bea}{\begin{eqnarray}}
\newcommand{\eea}{\end{eqnarray}}
\newcommand{\bean}{\begin{eqnarray*}}
\newcommand{\eean}{\end{eqnarray*}}

\def\({\left(}
\def\){\right)}
\def\[{\left[}
\def\]{\right]}

\def\frac#1#2{{#1 \over #2}}

\newcommand{\half}{\frac{1}{2}}

\renewcommand{\a}{\alpha}
\renewcommand{\b}{\beta}

\renewcommand{\r}{\rho}
\newcommand{\s}{\sigma}

\newcommand{\m}{\mu}
\newcommand{\n}{\nu}

\renewcommand{\S}{\Sigma}

\newcommand{\CS}{{\cal S}}

\newcommand{\lsim}{\,\raise.3ex\hbox{$<$\kern-.75em\lower1ex\hbox{$\sim$}}\,}
\newcommand{\gsim}{\,\raise.3ex\hbox{$>$\kern-.75em\lower1ex\hbox{$\sim$}}\,}

\def\II{\relax{I\kern-.10em I}}

\usepackage{color}
\usepackage{multirow}
\usepackage[super]{nth}

\definecolor{Blueberry}{rgb}{0.25,0,0.65}
\definecolor{Strawberry}{rgb}{0.65,0,0.25}
\definecolor{somefruit}{rgb}{0.0,.8,0.2}

\newcommand{\AdSd}{{$\mathrm{AdS}_{d+1}$}}
\newcommand{\AdSfour}{{$\mathrm{AdS}_{4}$}}
\newcommand{\AdS}{{$\mathrm{AdS}$}}

\newcommand{\ssection}[1]{\begin{center}{\em #1}\end{center} }

\begin{document}

\title{Dynamical Spacetimes from Numerical Hydrodynamics}
\author{Allan Adams}
\affiliation{Center for Theoretical Physics, Massachusetts Institute of Technology,  Cambridge, MA  02139}
\author{Nathan Benjamin}
\affiliation{Center for Theoretical Physics, Massachusetts Institute of Technology,  Cambridge, MA  02139}
\affiliation{Stanford Institute for Theoretical Physics, Stanford University,  Stanford, CA  94305}
\author{Arvin Moghaddam}
\affiliation{Center for Theoretical Physics, Massachusetts Institute of Technology,  Cambridge, MA  02139}
\affiliation{Center for Theoretical Physics, University of California, Berkeley,  Berkeley, CA 94720}
\author{Wojciech Musial}
\affiliation{Center for Theoretical Physics, Massachusetts Institute of Technology,  Cambridge, MA  02139}
\affiliation{Center for Theoretical Physics, University of California, Berkeley,  Berkeley, CA 94720}

\begin{abstract}\noindent
We numerically construct dynamical asymptotically-\AdSfour\ metrics by evaluating the fluid/gravity metric on numerical solutions of dissipative hydrodynamics in (2+1) dimensions.  The resulting numerical metrics satisfy Einstein's equations in (3+1) dimensions to high accuracy.
\end{abstract}

\preprint{MIT-CTP-4607}
\maketitle

%
%

Holography provides a precise relationship between black holes in \AdSd\ and QFTs in $d$-dimensions at finite density and temperature.  When the QFT state lies in a hydrodynamic regime (i.e. when $d$-dimensional gradients are {sufficiently small} that the stress-tensor may be expanded as a power-series in derivatives), the dual spacetime metric may also be so expanded, leading to an analytic ``fluid/gravity'' map between solutions of hydrodynamics in $d$ dimensions and dynamical asymptotically-\AdSd\ solutions of the Einstein equations.  

This 
suggests a simple strategy for constructing numerical solutions of the Einstein equations in asymptotically \AdSd\ spacetimes: 
rather than solve the ($d$+1)-dimensional Einstein equations numerically, we may numerically solve $d$-dimensional hydrodynamics and use the fluid/gravity map to analytically construct the corresponding metric. 
{So long as the fluid flow lies sufficiently deep in the hydro regime, the hydro equations can be truncated at a desired order in the gradient expansion; the resulting fluid/gravity metric is then an analytic solution of the Einstein equations up to errors of corresponding order.}  
Na\"ively, this should be a much easier calculation, 
since one need only solve PDEs in 3$d$, rather than 4$d$.  
The main question in principle is whether the resulting algorithm is sufficiently robust to small departures from the true hydro solution and numerical errors to be useful. 
In practice, this will also serve as an independent check of the analytic results in the literature.

In this paper we construct numerical solutions to \nth{2} order dissipative relativistic hydrodynamics, evaluate the fluid/gravity metric on the resulting flows, and test the satisfaction of the Einstein equations on the corresponding dynamical spacetimes. 
Implementing the fluid/gravity map revealed minor typos in the literature which we corrected by re-deriving the fluid/gravity analytic map.\footnote{We thank Jae-Hoon~Lee and Mark~Van~Raamsdonk for discussions on this point.} 
The resulting numerical metrics satisfy the Einstein equations with great precision.  
Concretely, we find that when the flow is hydrodynamic, the \nth{0} order fluid/gravity metric provides good approximation of a solution to the Einstein equations for a wide range of fluid flows within the hydro regime, and that the \nth{1} and \nth{2} order corrections improve the accuracy of this fluid metric appropriately, leading to accurate numerical metrics computed at low computational expense.

%
%
\vfill
\ssection{Review of Fluid/Gravity}

We begin by recalling the essential features of the fluid/gravity correspondence for a (2+1) fluid as given in \cite{VanRaamsdonk:2008fp,Bhattacharyya:2008jc,Bhattacharyya:2008mz,Hubeny:2011hd}.  
At equilibrium, a fluid moving with constant $3$-velocity $u^{\mu}$ at temperature $T$ is dual to an asymptotically \AdSfour\ black brane described by the metric \footnote{{Throughout this paper, $m, n \in\{0,..,3\}$ denote bulk 3+1 indices, $\m, \n \in \{0,1,2\}$ denote boundary 2+1 indices, while  $i, j \in \{1,2\}$ denote boundary spatial indices. }}
\be
ds^{2}  =  -2u_{\mu} dx^{\m}dr -r^{2}f(br)\,u_{\m} u_{\n}dx^{\m}dx^{\n} + r^{2} P_{\m\n} dx^{\m}dx^{\n}\,, \non
\ee
where $b = {3\over4\pi T}$ is the rescaled inverse temperature, ${P^{\m\n} \!=\! \eta^{\m\n}\!+\!u^{\m}u^{\n}}$  projects onto directions transverse to $u^{\m}$, and $f(\r)=1-{1\over\r^{3}}$ is the emblackening factor.  It is readily checked that this metric solves the Einstein equations so long as $u^{\m}$ and $T$ are constant.   

If $u^{\m}({\bf x})$ and $T({\bf x})$ vary in space and time, it 
follows that this metric is again a solution of the Einstein equations when expanded to leading (trivial) order in gradients.   
When all gradients are small,  it is possible to systematically improve the metric order-by-order in gradients to construct a solution of the ($d$+1)-dimensional Einstein equations provided that $u^{\m}({\bf x})$ and $T({\bf x})$ solve the equations of $d$-dimensional hydrodynamics.  

Demonstrating this is simplified by working in a gauge in which the $n^{\rm th}$-order corrections take the form,
\bea
&&\!\!\!\!\!\!
ds^{2}_{(n)} =
- h^{(n)} ( 2u_{\mu} dx^{\m}dr + r^{2} P_{\m\n} dx^{\m}dx^{\n} ) \non \\
&&
~+ k^{(n)}   u_{\m} u_{\n}dx^{\m}dx^{\n}    
+j^{(n)}_{\n}   {2\over r} u_{\m} dx^{\m}dx^{\n} 
+\a^{(n)}_{\m\n}  dx^{\m}dx^{\n}  \non .
\eea
Upon expanding to $n^{\rm th}$-order in gradients, the $rr$ and $\m\n$ components of the Einstein equation reduce to linear equations for the  functions $h^{(n)}$, $k^{(n)}$, $j^{(n)}$ and $\a^{(n)}$,
\bea
{1\over r^{4}}{d\over dr}\(r^{4}{d\over dr}h^{(n)}(r,{\bf x})\) &=& S_{h}^{(n)}\!(r,u^{\m}({\bf x})) \non \\
{d\over dr}\(-{2\over r}k^{(n)}(r,{\bf x}) + (1-4r^{3})h^{(n)}(r,{\bf x})\) &=& S_{k}^{(n)}\!(r,u^{\m}({\bf x})) \non \\
{r\over2}{d\over dr} \( {1\over r^{2}} {d\over dr} j_{\n}^{(n)}(r,{\bf x}) \) &=& S_{j_{\n}}^{(n)}\!(r,u^{\m}({\bf x})) \non \\
{d\over dr}\(-{1\over2}r^{4}f(r){d\over dr}\a_{\m\n}^{(n)}(r,{\bf x})\) &=& S_{\a_{\m\n}}^{(n)}\!(r,u^{\m}({\bf x}))\,, \non 
\eea
where the ``source'' functions $S^{(n)}_{*}\!(r,u^{\m}({\bf x}))$ may be explicitly computed order-by-order in the gradient expansion and depend only on $n$ or fewer derivatives. 
These equations are manifestly local in the fluid dimensions, $\bf x$, so solving them reduces to a series of 
numerically-simple 1-dimensional integration problems.  

At first order, the sources take simple forms, 
$S_{h}^{(1)}  =  0$,  
$S_{k}^{(1)}  =  -4r\nabla\!\cdot\!\b$,  
$S_{j_{i}}^{(1)}  =  -{1\over r}\partial_{t}\b_{i}$, and 
$S_{\a_{ij}}^{(1)}  =  2r\s_{ij}$,
where $\s_{\ij} = \half\( \nabla_{i}\b_{j} +\nabla_{j}\b_{i} -\half\delta_{ij}\nabla\!\cdot\!\b \)$, $\b_{i}$ is the fluid velocity in the local rest frame at position $\bf x$,  $\b_{i}({\bf x})=0$, and $i$, $j$ label the spatial dimensions.\footnote{Note that we have fixed to a local co-moving frame, which simplifies the calculations; it is straightforward to promote these relations to covariant expressions.}

Given these sources, the equations above can be integrated along the radial direction to give \footnote{Appropriate boundary conditions follow from regularity, renormalizability and our choice of gauge, as explained in \cite{Bhattacharyya:2008jc}.}\!: 
${h}^{(1)}  \!=\! 0 $, 
${k}^{(1)}  \!=\!  -{1\over 4} S_{k}^{(1)}$, 
${j^{(1)}_{\m}}  =  -r^{3} S_{j_{\m}}^{(1)}$, and 
${\a^{(1)}_{\m\n}} = {1\over r} S_{\a_{\m\n}}^{(1)} F(r) $, 
where 
$$
F(r)={-1\over\sqrt{3}}{\rm Tan}^{-1}\({2r+1\over\sqrt{3}}\)+\half\log\({1+r+r^{2}\over r^{2}}\) +{\sqrt{3}\pi\over6}\,,
$$
as presented in \cite{VanRaamsdonk:2008fp}.

At second order the sources are somewhat more cumbersome, so we refer the reader to \cite{VanRaamsdonk:2008fp}\ whose \nth{2} order sources we have analytically verified modulo minor typos in $S^{(2)}_{k}$ and $F_{2}$, which we find take the values,
\bea
&&~~~~~~S^{(2)}_{k}=2\CS_{3}+\half\CS_{5} -{1+4r^{3}\over 2r^{3}}\CS_{6} +F_{2}(r)\,\CS_{7}\,, \non\\
&&F_{2} = {2(1+r)(1-4r^{3})\over r(1+r+r^{2})}F(r) +{-1+2r+4r^{2}+4r^{3} \over r(1+r+r^{2})}\,. ~~~~\non
\eea

While the resulting metric satisfies the $rr$ and $\m\n$ components of the Einstein equations, the $r\mu$ components of the Einstein equations impose a further set of $d$ conditions.  
{These constraints are equivalent to the conservation of the fluid stress tensor built from $u^{\m}({\bf x})$ and $T({\bf x})$ at the corresponding order in the gradient expansion with a specific set of transport coefficients \cite{VanRaamsdonk:2008fp,Bhattacharyya:2008jc}.}
Thus, given a solution $u^{\mu}({\bf x})$ and $T({\bf x})$ of the hydro equations of motion, we can simply plug this flow into the fluid-gravity metric equations and, upon solving a set of 1-dimensional ODEs along the radial direction, construct a numerical metric which satisfies the full Einstein equations to the appropriate order in the fluid gradients.

%
%
\ssection{Numerical Methods}

Our calculation is naturally divided into two steps.  We first construct numerical solutions of the hydrodynamic equations; we then evaluate the fluid/gravity metric on these solution.  We estimate our errors by evaluating the Einstein equations on the resulting numerical metric.


Solving the relativistic hydrodynamic equations is by now relatively standard.  In principle, for constructing the \nth{2} order metric we need only solve relativistic hydro at \nth{1} order; 
in practice, however, \nth{1} order hydro is dynamically unstable.  We regulate this instability by solving the full \nth{2} order hydro, using for good measure the specific transport coefficients derived in the fluid/gravity analysis.  The corresponding stress tensor is given in \mbox{Eq (3)} of \cite{VanRaamsdonk:2008fp}.  Since we work with a \nth{2} order stress tensor, the resulting conservation equations are third order in time-derivatives of the fluid variables.  
To avoid spurious solutions 
and simplify our calculation, we treat the dissipative stress tensor $\Pi^{\m\n}$ as an independent dynamical variable whose evolution is determined by the constitutive relations a la Israel-Stewart. So long as we remain within the hydrodynamic regime, solutions of the resulting equations should relax toward solutions of the full equations.\footnote{{See for example \cite{Baier:2007ix}\ and references therein.}} We use the satisfaction of the constitutive relations as a check on our numerical solutions.

Upon fixing Landau gauge ($u_{\m}\Pi^{\m\n}$=0) and demanding conformal invariance ($\Pi^{\m}_{~\m}$=0), only 2 components of $\Pi^{\m\n}$ remain linearly independent.  A convenient parameterization for these two dissipative variables involves setting
\bea
\Pi_{xy} &=& (2+u_{x}^{2}+u_{y}^{2})\,\Pi\,, \non \\
\Pi_{xx} &=& 2(1+u_{x}^{2})\,\S \,+ 2u_{x}u_{y}\,\Pi \,. \non 
\eea
All other components of $\Pi_{\m\n}$ may then be determined in terms of regular functions of $u_{\mu}$ times $\Pi$ and $\S$.

The hydro equations then take the form $M(u) \dot{u} = b(u)$, 
where $u$ is a vector of our five variables ($u_{x}$, $u_{y}$, $T$, $\S$ and $\Pi$), while the $5\times5$ matrix $M(u)$ and the 5-vector $b(u)$ are non-linear functions of $u$ and its spatial (but not time) derivatives.  
To determine the time-derivative of our fields at a given point in space, we first evaluate the required spatial derivatives of our fields, compute $M$ and $b$, and then numerically solve this $5\times5$ matrix equation.  
Having computed the time-derivatives, 
we propagate the solution forward in time via standard techniques.

\newcommand{\LX}{1500}
\newcommand{\NX}{306}
\newcommand{\NZ}{32}

In our numerical computations we fix periodic boundary conditions in both spatial directions with period $L=\LX$, with the initial temperature set to a constant $T(0)={3\over4\pi}$.  We represent all fields pseudospectrally in a fourier basis of \NX$\times$\NX\ plane waves, computing spatial derivatives spectrally and propagating the system forward in time using Matlab's built-in general-purpose integrator, {\tt ode45}.
We focus on two classes of initial conditions: a superposition of the 1000 lowest-frequency plane waves with randomized amplitudes and phases; and, following \cite{Adams:2013vsa}, a line of 10 vorticity stripes generated by the velocity field $u_{y}=\cos(5 {2\pi\over L}x)$, to which we add small perturbations comprised of the five lowest-frequency plane waves with small random amplitudes and random phases.


We now turn to evaluating the fluid/gravity metric $g_{mn}({\bf x},r)$ on a solution $u^{\m}({\bf x})$ and $T({\bf x})$ of the hydro equations.  We represent the metric components spectrally.  
Along each boundary spatial dimension we again expand in a basis of \NX\ fourier modes.  Along the bulk radial dimension we work with the coordinate $z=1/r$ which extends from $z=0$ (the \AdS\ boundary) to $z=2$ (well inside the apparent horizon at $z=b$).  We expand our fields in a basis of \NZ\ Chebyshev modes along $z$.  Integration of the equations for $h^{(2)}$,  $k^{(2)}$ and $j^{(2)}_{i}$ is performed {analytically wherever possible} and otherwise spectrally, with the required boundary conditions imposed by subtraction.  The $\a^{(2)}_{ij}$ equation, however, must be solved as a 2-point boundary value problem due to the necessity of imposing one boundary condition (regularity) at the horizon and another at the boundary, at both of which points the equation is degenerate.  This is done via a single inversion of the boundary-blocked linear operator appearing in the $\alpha^{(2)}$ equation which is then multiplied against the bordered source $S^{(n)}_{\a}$.  The Einstein {tensor} is then evaluated on the resulting metric in the same spectral basis, with time derivatives of relevant GR tensors computed using a \nth{4} order finite differencing scheme.

Given an exact solution,  $E_{mn}\equiv G_{mn}-\half\Lambda g_{mn}$ would vanish point-wise.  We thus use $E_{mn}$ evaluated on the fluid metric as a measure of the accuracy of our solution.  More precisely, we estimate the maximum local error,
$$
{\cal E}_{loc}(t) =\, \max_{V,\mu,\nu} \, |E^{\mu \nu}|\,,
$$
where $V$ is the spatial computational domain, as well as the global RMS error,
$$
\bar{\cal E}(t)^{2} = { \int_{V} \sqrt{g} \,\,\(E^{m}_{~m}\)^{2}  \over  \int_{V} \sqrt{g} }\,.
$$
The value of these observables represents an estimate of the error in our numerical metric.  Note that these quantities scale as two powers of space-time gradients at leading order in the gradient expansion for the metric.

%
%
\begin{figure}[t!]
\includegraphics[scale = 0.9]{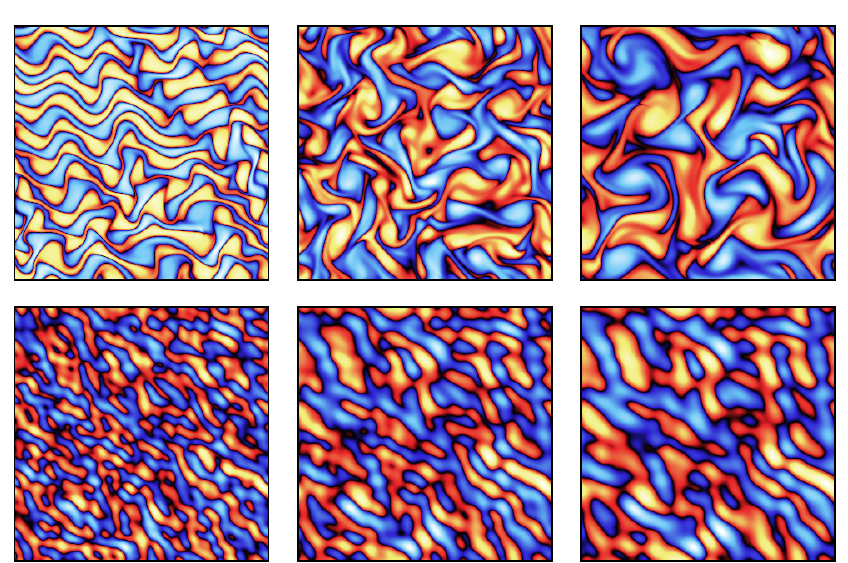}
\caption{Snapshots of the vorticity along the flow for both classes of initial conditions at time 1000, 2000 and 3000.  Top: weakly-perturbed vorticity stripe initial conditions rapidly decay via a classic two-stream instability into a slowly-relaxing quasi-turbulent state.  Bottom: random-wave initial conditions gradually fade away.  Red indicates positive vorticity, blue negative and black zero.} 
\end{figure}

\ssection{Results and Conclusions}

{Figure 1 displays the vorticity of the fluid at three moments along the numerical evolution of two typical fluid flows, one from each class of initial conditions.}  The resulting flows behave much as one expects of a normal 2d fluid in the hydro regime, with visible transfer of spectral weight from high to low wave numbers in a classic inverse cascade.

Figures 2 and 3 plot ${\cal E}_{loc}(t)$ and $\bar{\cal E}(t)$ respectively as a function of time for the \nth{0}, \nth{1} and \nth{2} order metrics.
On the left are the results for the random-wave initial condition; on the right, the vorticity stripe.  
As is apparent, the \nth{0} order metric is already a good approximation,  the \nth{1} order metric further reduces this error considerably, with the \nth{2} order corrections giving us an extremely accurate numerical solution of the bulk Einstein equations.  At the bottom of each plot is a measure of the satisfaction of the constitutive relations.  Errors in the constitutive relations are strongly correlated with errors in the bulk metric, as they should: the constitutive relations arise from the asymptotic radial constraint equations of the bulk Einstein equations.

\begin{figure*}[t!]
\includegraphics[scale = 0.55]{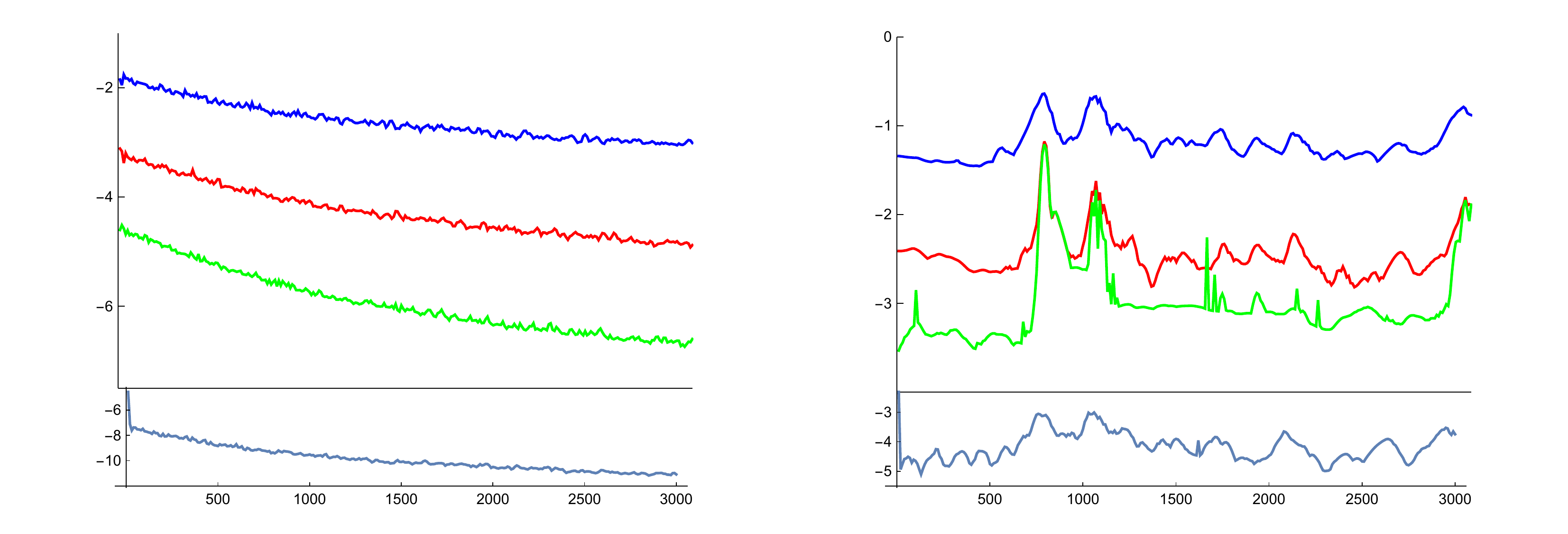}
\caption{Log base 10 error of the \nth{0} (blue), \nth{1} (red) and \nth{2} (green) order fluid/gravity metrics as measured by the maximum absolute value of the Einstein equation as a function of time.  Left: random-wave initial conditions.  Right: weakly-perturbed vorticity stripe initial conditions. The accuracy of the metric is excellent in both cases, converging as expected at subsequent orders. Note that the large increase in error around times 700 and 1000 for the bottom graph correspond to moments where the fluid gradients become unusually large.  The bottom plots indicate the convergence of the constitutive relations as a function of time along the corresponding flow.}
\end{figure*}

\begin{figure*}[t!]
\includegraphics[scale = 0.55]{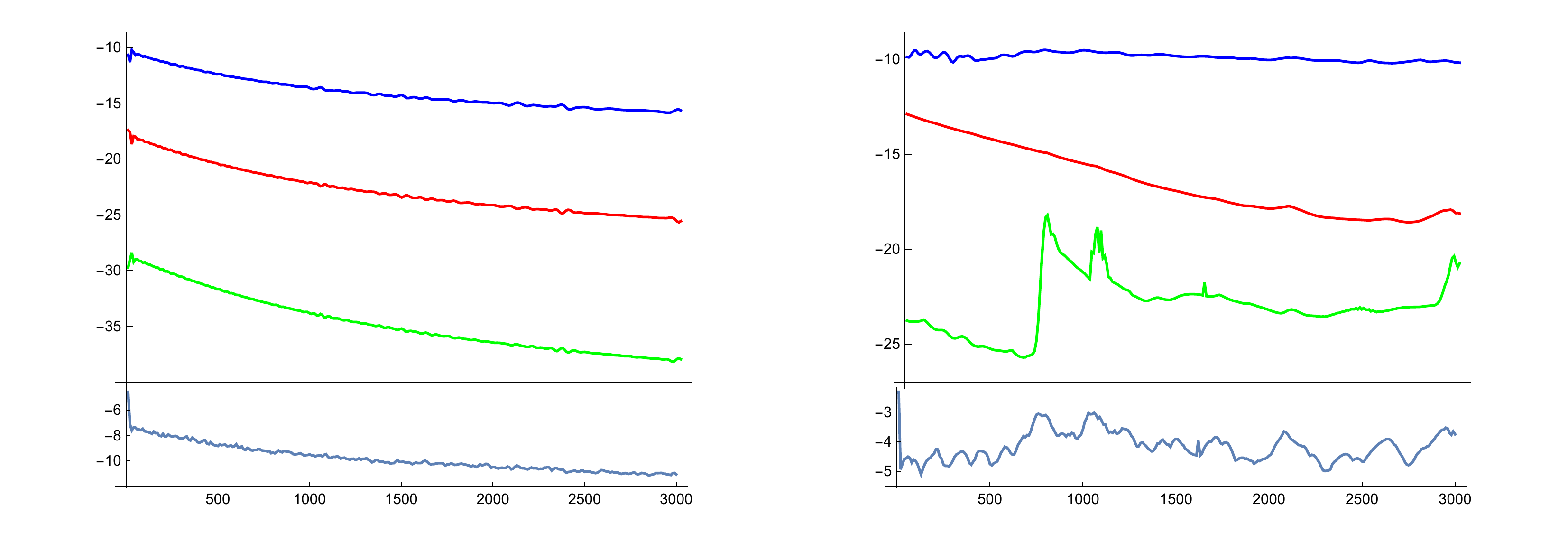}
\caption{Log base 10 error of the \nth{0} (blue), \nth{1} (red) and \nth{2} (green) order fluid/gravity metrics as measured by the average RMS value of the Einstein equation as a function of time.  Left: random-wave initial conditions.  Right: weakly-perturbed vorticity stripe initial conditions.}
\end{figure*}

%
To summarize, the fluid/gravity correspondence provides a novel, robust and fast algorithm for constructing non-equilibrium asymptotically-\AdS\ numerical solutions of the Einstein equations.  As we have seen, this approach is quantitatively effective even when the dynamics drive the system nonlinear and turbulent, so long as typical gradients remain bounded so that the hydrodynamic expansion is reliable.  To go beyond the hydro regime, or to test the fluid/gravity correspondence as one approaches the hydro regime, requires direct numerical solution of the Einstein equations, as in \cite{Chesler:2008hg,Chesler:2010bi,Adams:2013vsa}.  Nonetheless, the ease and efficiency of this approach makes it a useful tool for gravitational questions within the hydro regime. 


\vspace{5pt}

%
%

\ssection{Acknowledgments}

We thank Lincoln Carr, Paul Chesler, Ethan Dyer, Jae Hoon Lee, Luis Lehner, Hong Liu, R. Loganayagam and Mark Van Raamsdonk for helpful discussions. 
AA thanks the organizers of the ``Cosmology and Complexity 2012'' and ``Relativistic hydrodynamics and the gauge-gravity duality at the Technion'' workshops where this work has been discussed, and the Aspen Center for Physics for hospitality.
The work of AA is supported by the U.S. Department of Energy under grant Contract Number  DE-SC00012567.  The work of NB, AM and WM was supported by MIT's Undergraduate Research Opportunities Program.

\bibliographystyle{hunsrt}
\bibliography{FluidGravity}
\end{document}